\documentclass[conference]{IEEEtran}
\IEEEoverridecommandlockouts
\usepackage{cite}
\usepackage{amsmath,amssymb,amsfonts}
\usepackage{algorithmic}
\usepackage{graphicx}
\usepackage{textcomp}
\usepackage{xcolor}
\usepackage[pass]{geometry}
\usepackage{pdfpages}
\def\BibTeX{{\rm B\kern-.05em{\sc i\kern-.025em b}\kern-.08em
    T\kern-.1667em\lower.7ex\hbox{E}\kern-.125emX}}
    
\begin{document}

\title{Simulating Quantum Systems with NWQ-Sim on HPC\\
}

\author{\IEEEauthorblockN{In-Saeng Suh}
\IEEEauthorblockA{\textit{National Center for Computational Sciences} \\
\textit{Oak Ridge National Laboratory}\\
Oak Ridge, Tennessee, USA \\
suhi@ornl.gov} 
\and
\IEEEauthorblockN{Ang Li}
\IEEEauthorblockA{\textit{Physical and Computational Sciences Directorate} \\
\textit{Pacific Northwest National Laboratory}\\
Richland, Washington, USA \\
ang.li@pnnl.gov}
}

\maketitle

\begin{abstract}
NWQ-Sim is a cutting-edge quantum system simulation environment designed to run on classical multi-node, multi-CPU/GPU heterogeneous HPC systems. 
In this work, we provide a brief overview of NWQ-Sim and its implementation in simulating quantum circuit applications, such as the transverse field Ising model. We also demonstrate how NWQ-Sim can be used to examine the effects of errors that occur on real quantum devices, using a combined device noise model. Moreover, NWQ-Sim is particularly well-suited for implementing variational quantum algorithms where circuits are dynamically generated. Therefore, we also illustrate this with the variational quantum eigensolver (VQE) for the Ising model. In both cases, NWQ-Sim's performance is comparable to or better than alternative simulators. We conclude that NWQ-Sim is a useful and flexible tool for simulating quantum circuits and algorithms, with performance advantages and noise-aware simulation capabilities.

\end{abstract}


\section{Introduction}

In recent years, Noisy-Intermediate-Scale-Quantum (NISQ) \cite{JohnPreskill2018} based quantum computing (QC) has made significant progress \cite{FrankArute2019,SergioBoixo2018}. However, despite these advances, the wider QC community still heavily relies on classical machine simulations for the development and validation of quantum algorithms. This is particularly true for the promising quantum variational algorithms (VQA) \cite{QC-online}, including variational quantum eigensolvers (VQE) \cite{AbhinavKandala2017, JonathanRomero2018}, quantum neural networks (QNN) \cite{KerstinBeer2020, stein2022quclassi}, and quantum approximated optimization algorithms (QAOA) \cite{EdwardFarhi2014, GianGiacomoGuerreschi2019}.

There are several reasons for this dependence on classical simulations. Firstly, the limited availability of quantum computing resources such as those provided by IBMQ or Azure Quantum means that a large number of users have to share them, resulting in long waiting times for quantum resource allocation. 
Secondly, since parameterized variational circuits are typically deep,
as a result, the depth of these circuits can easily exceed the maximum gate allowance of NISQ devices due to their short coherence time. 
Finally, current quantum computers are not quantum-error-correction (QEC) protected \cite{RamiBarends2014, DanielALidar2013}, resulting in high error rates. Consequently, simulations are necessary for validating quantum algorithms and debugging circuits \cite{JonathanCarter2017, TysonJones2019}.


A variety of quantum circuit simulators on classical computers have been proposed \cite{ListQC}. However, the majority of these simulators focus on logical qubits within an ideal isolated system setting, where complete gate fidelity can be guaranteed, rather than physical qubits within a practical, open environment that is subject to inevitable noise. 
Designing an efficient and scalable QC simulator on classical high-performance computing (HPC) systems presents significant challenges. In the context of today's heterogeneous supercomputers, the performance of an application largely depends on how effectively accelerators, such as GPUs, are utilized. However, users still desire a unified and easy-to-use programming interface that is suitable for their domains.

In this context, we introduce the NWQ-Sim simulator which overcomes other simulator's challenges and its implementation in simulating quantum circuits and VQE for the time-dependent transverse field Ising model. We also use a device noise model to examine the effects of errors that occur on real quantum devices. Finally, we compare the performance of the NWQ-Sim and IBM Qiskit simulators.

\section{NWQ-Sim simulator}

\begin{figure}[!t]
    \centering
    \vspace{-10pt}
\includegraphics[scale=0.28]{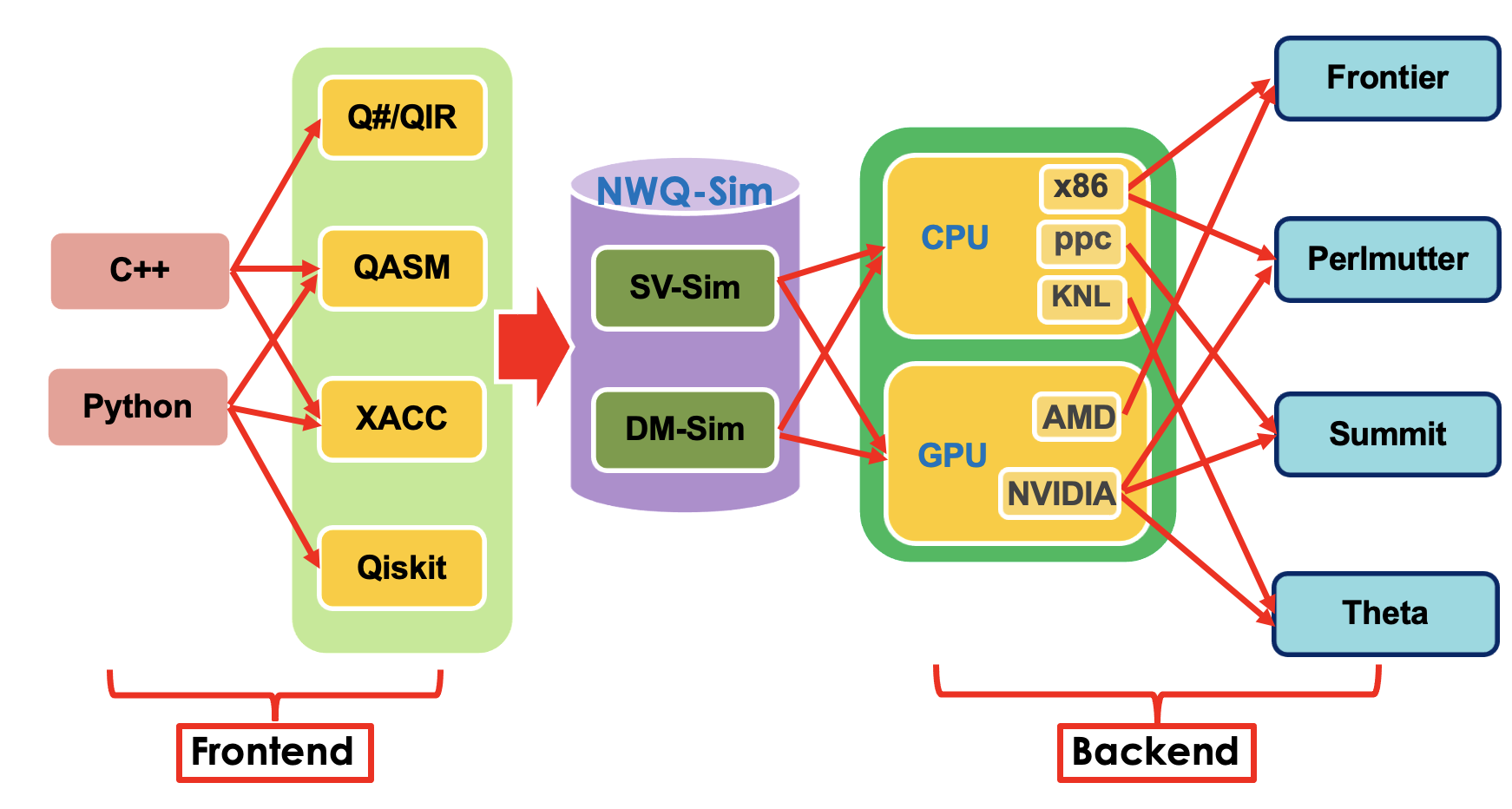}
    \vspace{-10pt}
\caption{The framework of the NWQ-Sim simulator}
\label{fig:nwq-sim}
\end{figure}

NWQ-Sim is a quantum system simulation environment that runs on classical multi-node, multi-CPU/GPU heterogeneous HPC systems \cite{angli2021sv,angli2020dm}.
It combined two different types of simulators - a state vector simulator SV-Sim for high performance quantum simulation and a density matrix simulator DM-Sim for noise-aware simulation. It supports C++, Python, Q\#/QIR, Qiskit, QASM, XACC~\cite{mccaskey2020xacc} as the frontends, 
and x86/PPC CPU, NVIDIA/AMD GPU as backends. Fig. \ref{fig:nwq-sim} shows the NWQ-Sim framework.

\begin{figure}[!t]
    \centering
     \vspace{-0.5cm}
\includegraphics[scale=0.40]{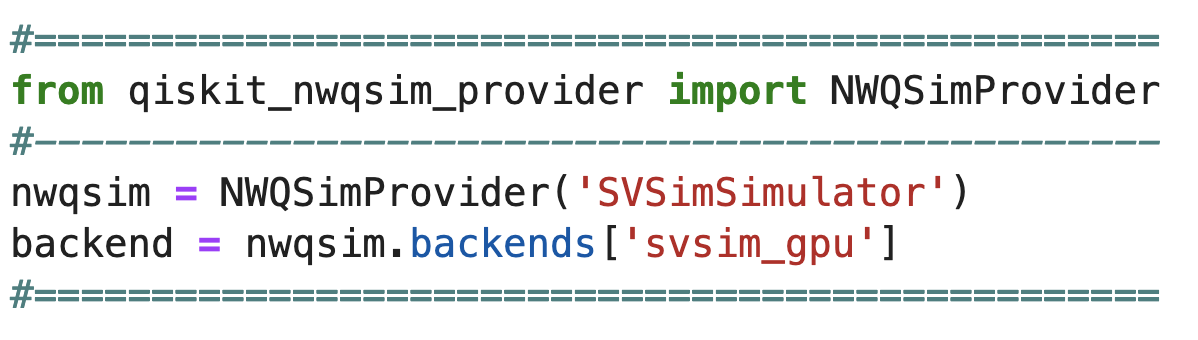}
    \vspace{-0.5cm}
\caption{Importing the SV-Sim simulator from NWQ-Sim}
\label{fig:nwq-svsim}
\end{figure}

To run a quantum circuit on NWQ-Sim, first one will need to import \textsf{NWQSimProvider} as shown in Fig. \ref{fig:nwq-svsim} and then specify the backend to use. The backend corresponds to the simulator that we want to use and is defined by the provider. 

\section{Quantum simulations with a noisy model}

As a first use case, a simulation was employed to model the magnetization and temporal variations of a spin chain governed by Ising-type interactions. 
Here, we will use these precise outcomes as a reference point for our quantum simulation of the Transverse Field Ising Model (TFIM). For simple experiments, let's consider $n=4$ spin chain. 
We optimize the quantum circuit and implement it on IBM Quantum Qiskit and NWQ-Sim simulators.

\begin{figure}[t!]
\centering
\includegraphics[scale=0.35]{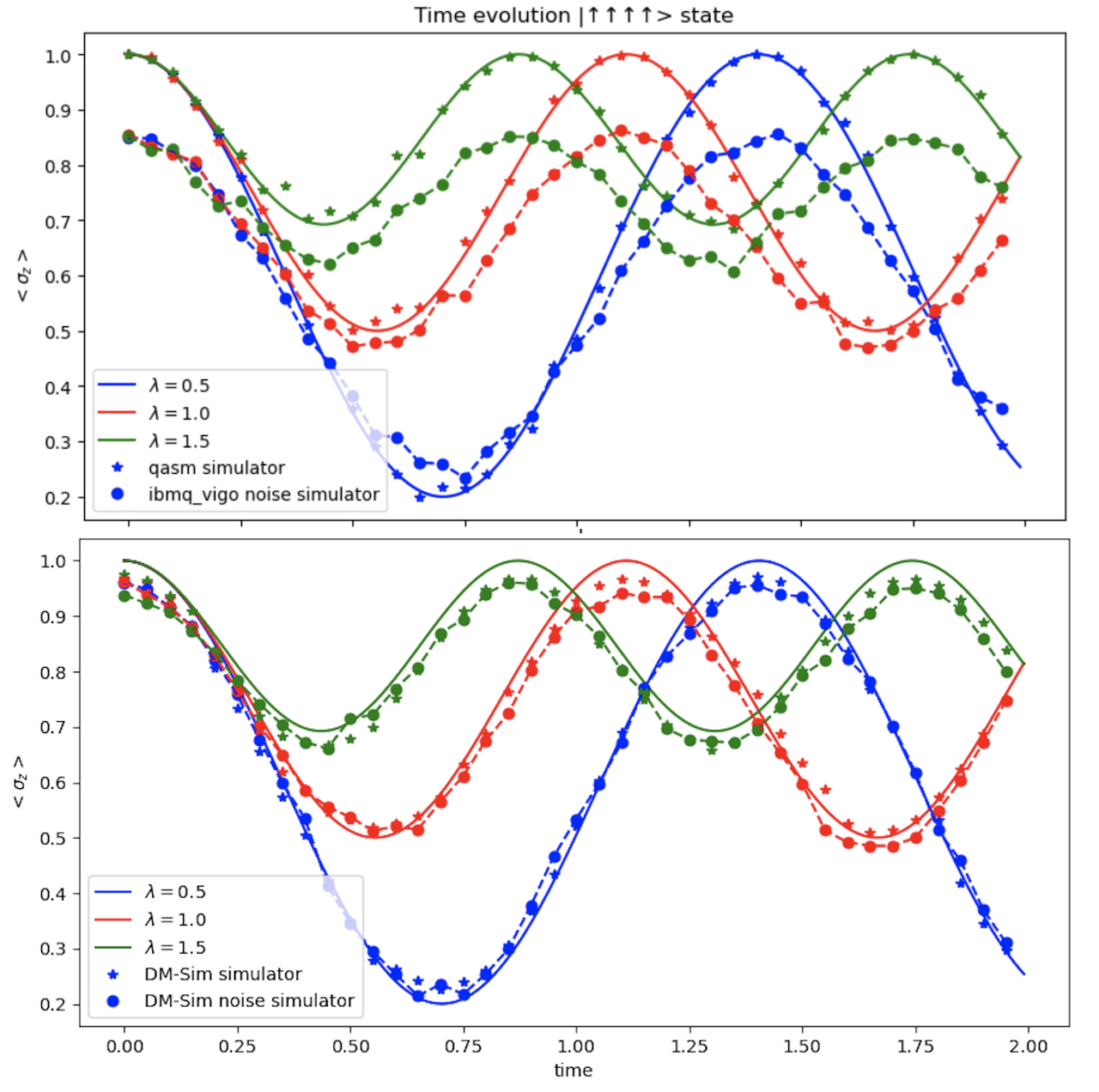}
\caption{Time evolution simulation of transverse magnetization, $\langle\sigma_z\rangle$, for the state $|\uparrow\uparrow\uparrow\uparrow\rangle$ of a $n=4$ Ising spin chain. Solid line represents the exact result in comparison with the experimental simulations on Qiskit Aer and NWQ-Sim noisy simulators represented by scatter points.}
\vspace*{-0.6cm}
\label{Fig:Fig-3}
\end{figure}

Fig.~\ref{Fig:Fig-3} shows the time evolution of transverse magnetization $\langle \sigma_z \rangle$. 
We also use the device backend noise model to do noisy simulations of the quantum circuits to investigate the effects of errors which occur on real quantum devices in both Qiskit Aer and NWQ-Sim. The noise model includes depolarization noise, thermal relaxation noise, and measurement noise. 
We use real noise data for an IBM Quantum device using the data for ibmq$\_$vigo stored in Qiskit Terra \cite{qiskit21a}. 

\begin{figure}[t!]
\centering
\includegraphics[scale=0.38]{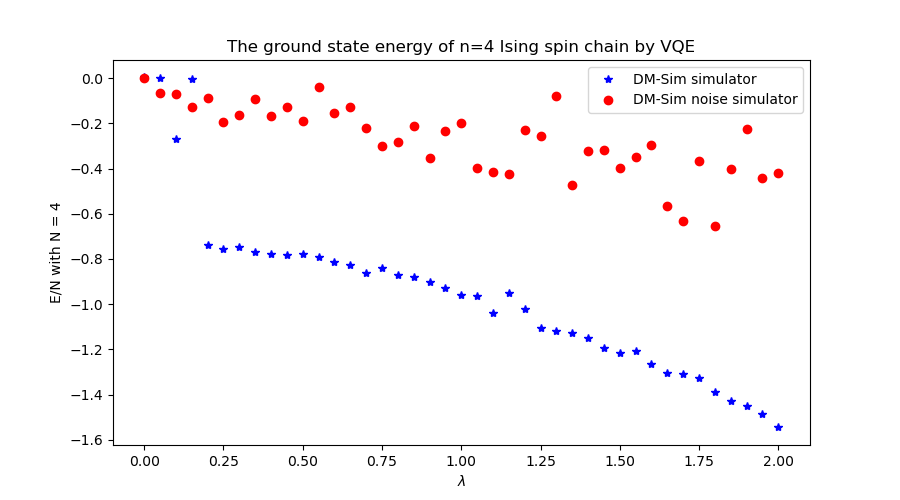}
\caption{VQE simulation in no-noise (star) and noise 
 (solid circle) with NWQ-Sim}
\vspace*{-0.1cm}
\label{Fig:Fig-4-vqe}
\end{figure}

\begin{figure}[t!]
\centering
\includegraphics[scale=0.38]{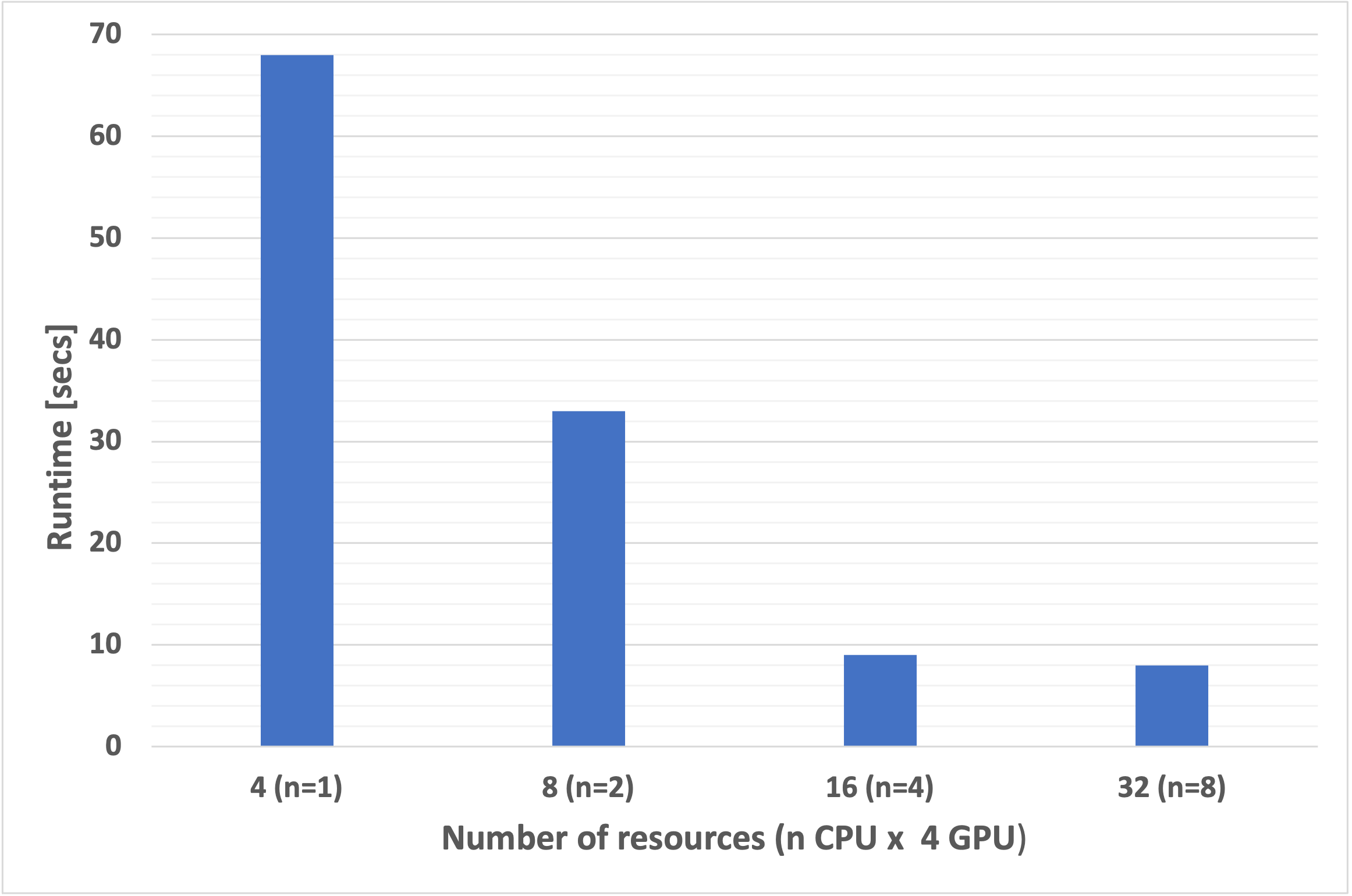}
\caption{QASMBench \cite{10.1145/3550488} with NWQ-Sim on OLCF Summit}
\vspace*{-0.6cm}
\label{Fig:Fig-qasmbench-summit}
\end{figure}

In the VQE simulation experiments (Fig. \ref{Fig:Fig-4-vqe}), we use the \textrm{COBYLA} method as a minimizer \cite{LohitPotnuru2021}. By comparing the noise VQE simulations with IBM Qiskit and NWQ-Sim simulators, we can further investigate the effect of noise on real quantum computers.
Fig. \ref{Fig:Fig-qasmbench-summit} shows QASM benchmark result for NWQ-Sim simulator on the Summit supercomputer.
We perform all simulations on the Frontier \cite{Frontier} and Summit supercomputer at ORNL \cite{Summit} and the Perlmutter supercomputer at NERSC \cite{Perlmutter}.

\section{Conclusion}

NWQ-Sim is a state-of-the-art quantum system simulation environment that offers advanced features for simulating quantum circuits and algorithms on classical multi-node, multi-CPU/GPU heterogeneous HPC systems. Its versatile simulation infrastructure, including the state vector simulator (SV-Sim) and the density matrix simulator (DM-Sim), provides high-performance quantum simulation as well as noise-aware simulation. NWQ-Sim's efficient inter-communication management capabilities make it particularly suitable for large-scale simulations.

In this work, we have provided an overview of NWQ-Sim and its implementation in simulating quantum circuit applications, showcasing its ability to examine the effects of errors on real quantum computers using a device noise model. For deep quantum circuit simulations, NWQ-Sim simulator shows comparable or better performance to the IBM Qiskit simulator under specific environments. Furthermore, NWQ-Sim's performance in implementing variational quantum algorithms, such as the variational quantum eigensolver (VQE) for the Ising model, gives better performance than other simulators. For example, it shows about $50\%$ faster on the GPU-based supercomputer.

\section*{Acknowledgment}
The research was supported by the Exascale Computing Project (17-SC-20-SC), a collaborative effort of the U.S. Department of Energy Office of Science and the National Nuclear Security Administration. This material is based upon work supported by the U.S. Department of Energy, Office of Science, National Quantum Information Science Research Centers, Quantum Science Center (QSC). This research used resources of the Oak Ridge Leadership Computing Facility at the Oak Ridge National Laboratory, which is supported by the Office of Science of the U.S. Department of Energy under Contract No. DE-AC05-00OR22725. This research used resources of the National Energy Research Scientific Computing Center (NERSC), a U.S. Department of Energy Office of Science User Facility located at Lawrence Berkeley National Laboratory, operated under Contract No. DE-AC02-05CH11231 using Awards ERCAP0023224 and ERCAP0023053. \\
{\it Notice}: This manuscript has been authored by UT-Battelle, LLC, under contract DE-AC05-00OR22725 with the US Department of Energy (DOE). The US government retains and the publisher, by accepting the article for publication, acknowledges that the US government retains a nonexclusive, paid-up, irrevocable, worldwide license to publish or reproduce the published form of this manuscript, or allow others to do so, for US government purposes. DOE will provide public access to these results of federally sponsored research in accordance with the DOE Public Access Plan (https://www.energy.gov/doe-public-access-plan).


\begin{thebibliography}{00}
\bibitem{JohnPreskill2018} John Preskill. Quantum computing in the nisq era and beyond. Quantum, 2:79, 2018. 
\bibitem{FrankArute2019} Frank Arute, Kunal Arya, Ryan Babbush, Dave Bacon, Joseph C Bardin, Rami Barends, Rupak Biswas, Sergio Boixo, Fernando GSL Brandao, David A Buell, et al. Quantum supremacy using a programmable superconducting processor. Nature, 574(7779):505–510, 2019.
\bibitem{SergioBoixo2018} Sergio Boixo, Sergei V Isakov, Vadim N Smelyanskiy, Ryan Babbush, Nan Ding, Zhang Jiang, Michael J Bremner, John M Martinis, and Hartmut Neven. Characterizing quantum supremacy in near-term devices. Nature Physics, 14(6):595–600, 2018.
\bibitem{QC-online} Quantum computing: Qubit and entanglement. Review of the Universe: Structures, Evolutiuons, Observations, and Theories.
\bibitem{AbhinavKandala2017} Abhinav Kandala, Antonio Mezzacapo, Kristan Temme, Maika Takita, Markus Brink, Jerry M Chow, and Jay M Gambetta. Hardware-efficient variational quantum eigensolver for small molecules and quantum magnets. nature, 549(7671):242–246, 2017.
\bibitem{JonathanRomero2018} Jonathan Romero, Ryan Babbush, Jarrod R McClean, Cornelius Hempel, Peter J Love, and Alán Aspuru-Guzik. Strategies for quantum computing molecular energies using the unitary coupled cluster ansatz. Quantum Science and Technology, 4(1):014008, 2018.
\bibitem{KerstinBeer2020} Jonathan Romero, Ryan Babbush, Jarrod R McClean, Cornelius Hempel, Peter J Love, and Alán Aspuru-Guzik. Strategies for quantum computing molecular energies using the unitary coupled cluster ansatz. Quantum Science and Technology, 4(1):014008, 2018.
\bibitem{stein2022quclassi} Samuel A Stein, Betis Baheri, Daniel Chen, Ying Mao, Qiang Guan, Ang Li, Shuai Xu, and Caiwen Ding. Quclassi: A hybrid deep neural network architecture based on quantum state fidelity. Proceedings of Machine Learning and Systems, 4:251–264, 2022.
\bibitem{EdwardFarhi2014} Edward Farhi, Jeffrey Goldstone, and Sam Gutmann. A quantum approximate optimization algorithm. arXiv preprint arXiv:1411.4028, 2014.
\bibitem{GianGiacomoGuerreschi2019} Gian Giacomo Guerreschi and Anne Y Matsuura. Qaoa for max-cut requires hundreds of qubits for quantum speed-up. Scientific reports, 9:1–7, 2019.
\bibitem{RamiBarends2014} Rami Barends, Julian Kelly, Anthony Megrant, Andrzej Veitia, Daniel Sank, Evan Jeffrey, Ted C White, Josh Mutus, Austin G Fowler, Brooks Campbell, et al. Superconducting quantum circuits at the surface code threshold for fault tolerance. Nature, 508(7497):500–503, 2014.
\bibitem{DanielALidar2013} Daniel A Lidar and Todd A Brun. Quantum error correction. In Quantum error correction. Cambridge university press, 2013.
\bibitem{JonathanCarter2017} Jonathan Carter, David Dean, Greg Hebner, Jungsang Kim, Andrew Landahl, Peter Maunz, Raphael Pooser, Irfan Siddiqi, and Jeffrey Vetter. Ascr report on a quantum computing testbed for science. Technical report, USDOE Office of Science (SC), Washington, DC (United States), Advanced Scientific Computing Research Program, 2017.
\bibitem{TysonJones2019} Jones, Tyson and Brown, Anna and Bush, Ian and Benjamin, Simon C, QuEST and high performance simulation of quantum computers, Scientific reports, 9, 1 (2019)
\bibitem{ListQC} List of QC simulators. {\rm https://www.quantiki.org/wiki/list\-qc\-simulators}.
\bibitem{angli2021sv} Ang Li and Sriram Krishnamoorthy. SV-Sim: Scalable pgas-based state vector simulation of quantum circuits. In Proceedings of the International Conference for High Performance Computing, Networking, Storage and Analysis, 2021.
\bibitem{angli2020dm} Ang Li, Omer Subasi, Xiu Yang, and Sriram Krishnamoorthy. Density matrix quantum circuit simulation via the bsp machine on modern gpu clusters. In Proceedings of the International Conference for High Performance Computing, Networking, Storage and Analysis, 2020.
\bibitem{mccaskey2020xacc} Alexander J McCaskey, Dmitry I Lyakh, Eugene F Dumitrescu, Sarah S Powers, and Travis S Humble. Xacc: a system-level software infrastructure for heterogeneous quantum–classical computing. Quantum Science and Technology, 5(2):024002, 2020.
\bibitem{qiskit21a} Qiskit: An open-source framework for quantum computing. {\rm https://qiskit.org}. 
\bibitem{LohitPotnuru2021} Lohit Potnuru. Finding the ground state of the transverse Ising model.  {\rm https://github.com/LohitPotnuru/TransverseIsingModelQiskit.git}, 2021.
\bibitem{Frontier} Frontier at ORNL.  {\rm https://www.olcf.ornl.gov/frontier/}.    
\bibitem{Summit} Summit at ORNL.  {\rm https://www.olcf.ornl.gov/summit/}.  
\bibitem{Perlmutter} Perlmutter at NERSC. {\rm https://docs.nersc.gov/systems/perlmutter/}.
\bibitem{10.1145/3550488} Ang Li, Samuel Stein, Sriram Krishnamoorthy, and James Ang. {\rm QASMBench: A low-level quantum benchmark suite for nisq evaluation and simulation. ACM Transactions on Quantum Computing, 4(2), feb 2023.}
\end{thebibliography}


\end{document}